\documentclass[fleqn,twoside]{article}
\usepackage{espcrc2}

\usepackage{graphicx}
\usepackage[figuresright]{rotating}

\newcommand{\beq}{\begin{eqnarray}}
\newcommand{\eeq}{\end{eqnarray}}

\newcommand{\AmS}{{\protect\the\textfont2
  A\kern-.1667em\lower.5ex\hbox{M}\kern-.125emS}}

\title{Power Corrections to Perturbative QCD and OPE in Gluon Green Functions.}
\author{D. Becirevic\address[R1]{Dip di Fisica, Univ. di Roma ``La Sapienza and
INFN, Sezione di Roma,\\ P.le Aldo Moro 2, I-00185 Rome, Italy},
 Ph. Boucaud\address[LPT]{ Laboratoire de Physique
Th\'eorique, Unit\'e Mixte 
de Recherche du CNRS - UMR 8627\\
Universit\'e de Paris XI, B\^atiment 211, 91405 Orsay Cedex,
France}, F. De Soto\address[S]{Dpto. 
de F\'{\i}sica At\'omica, Molecular y Nuclear \\
Universidad de Sevilla, Apdo. 1065, 41080 Sevilla, Spain}, A. Le Yaouanc\addressmark[LPT], J.P. Leroy\addressmark[LPT], 
J. Micheli\addressmark[LPT],\\ O. P\`ene\addressmark[LPT], J Rodr\'\i
guez-Quintero\address[H]{Dpto. de F\'{\i}sica Aplicada \\
E.P.S. La R\'abida, Universidad de Huelva, 21819 Palos de la fra., Spain}, 
C. Roiesnel\address[X]{Centre de Physique Th\'eorique de l'\'Ecole Polytechnique\\
Unit\'e Mixte de Recherche C7644 du Centre National de 
la Recherche Scientifique\\
91128 Palaiseau Cedex, France }
}
       
\begin{document}

\begin{abstract}
We show that QCD Green functions in Landau Gauge exhibit 
sizable $O(1/\mu^2)$ corrections to the expected perturbative behavior 
at energies as high as 10 GeV.  We argue that these are due to a 
$\langle A^2 \rangle$-condensate which does not vanish in Landau gauge.
\vspace{1pc}
\end{abstract}

\maketitle

\section{Computing Green functions}

\subsection{The lattice settings}

We have performed a series of run in pure Yang-Mills SU3
with $\beta=6.0, 6.2, 6.4, 6.8$ i.e. $a^{-1}$ ranging from
0.03 to 0.1 fm. The lattices were hypercubic:  $16^4$, $24^4$
and once $32^4$. The statistics was of 1000 configurations per 
run. More details are available in \cite{phi,damir}.

\subsection{The method}
The method used to compute the strong coupling constant is described in
\cite{alles}-\cite{phi}. It consists in computing non-perturbatively the
renormalized  Green functions in MOM schemes. For a given scale $\mu$ the gluon
fields  are renormalized so that the gluon propagator at that scale is equal to
$1/\mu^2$ (times the standard tensor in Landau gauge). The renormalized coupling
constant is {\it defined} so that the three-gluon vertex function, projected on
the tree-level tensor, at that scale, is equal to the renormalized coupling
constant.  

In fact two different MOM schemes have been used. In the ``symmetric'' one, the
three-gluon vertex function is taken with gluon momenta
$p_1^2=p_2^2=p_3^2=\mu^2$, while the ``asymmetric'' one, called $\widetilde {\rm
MOM}$ corresponds to $p_1^2=p_2^2=\mu^2, p_3=0$

\subsection{Results without power corrections}

The three-gluon vertex function as defined above gives directly  a
nonperturbative value for $\alpha_S(\mu)$ for all  values of $\mu$ such that
$\mu^2=\sum_\nu p_\nu^2$. The gluon propagator evolves at large momenta according
to the perturbative formula. This evolution depends of course on the value of
$\alpha_s(\mu)$ at some  properly chosen scale $\mu$ (large enough to be in the
perturbative regime) in the range covered by the lattice calculation
\cite{damir}. In this subsection we neglect all power corrections. Thus we assume
that, for large enough $\mu$ the evolution of the propagator is fully described
by the perturbative formula taken to three loops.  Once $\alpha_s$ is known we
compute $\Lambda_{\rm QCD}$ to three loops in the considered scheme, and we
eventually translate it to the standard $\Lambda_{\overline {\rm MS}}$. Of
course, if we really are in the perturbative regime $\Lambda_{\overline {\rm
MS}}$ must be constant as the energy varies.  

We obtain from the three-point Green functions
{\small
\[
\alpha^{\widetilde{\rm MOM}}_S(4.3 \ {\rm GeV}) \ = 0.269(3);\,
\Lambda_{\overline {\rm MS}}(4.3 \ {\rm GeV}) \ = \ 299(7)
\label{three-alpha}
\]\[
\alpha^{\widetilde{\rm MOM}}_S(9.6 \ {\rm GeV}) \ = 0.176(2);\,
\Lambda_{\overline {\rm MS}}(9.6 \ {\rm GeV}) \ = \ 266(7)
\]
and from the two point-Green function
\[
\label{two-alpha}
\alpha^{\widetilde{\rm MOM}}_S(9.6 \ {\rm GeV}) \ = 0.193(3);\,
\Lambda_{\overline {\rm MS}}(9.6 \ {\rm GeV}) \ = \ 319(14)
\]}
We see that there is a problem in that $\Lambda_{\overline {\rm MS}}$
varies much more than allowed by the statistical errors.

\section{Power corrections.}

\begin{figure}[htb]
\includegraphics[width=15pc]{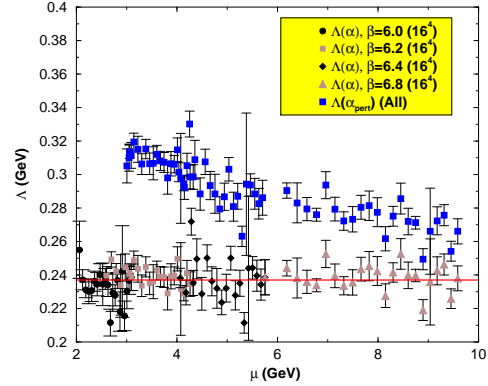}
\vspace*{-1.cm} 
\caption{The upper plot shows $\Lambda_{\overline {\rm MS}}$
computed without power corrections while the lower plot shows
$\Lambda_{\overline {\rm MS}}$ computed form the perturbative contribution
to $\alpha_s$ once the power correction has been substracted, eq. (\ref{power})
.}
\label{fig:lambda}
\end{figure}

We now add in our fits  $1/\mu^2$ power corrections to the three loop formula. In
\cite{direnzo} the strong coupling constant is fitted according to
\beq
\alpha_s(\mu^2)=\alpha_s(p^2)_{\rm 3 loops} + \frac {c} {\mu^2}
\label{power}\eeq
in a large energy window: 2 to 10 GeV. The result of the fit is 
\beq \Lambda_{\overline {\rm MS}} =  237 \pm 4\pm 10 {\rm MeV};\, c=0.63 
\mp 0.03 {\rm GeV}^2 \eeq
in strikingly good agreement with the result from the ALPHA collaboration 
\cite{Lusch} who has used a totally different method. The effect of power
corrections on the expected constancy of $\Lambda_{\overline {\rm MS}}$ is
illustrated in
fig \ref{fig:lambda}

\subsection{Power corrections and OPE}
What can be the theoretical reason for these large power corrections ? In the
theoretical framework of OPE a quadratic power requires  a condensate of a
dimension-two operator.  Since we are working in the Landau gauge, such an
operator exists:  $<A_\mu A^\mu>$. Interestingly enough it is the only dimension
two candidate for a vacuum expectation value in  the Landau gauge. 

We start from the OPE expansion of both the gluon propagator \cite{OPE} and the
symmetric three-gluon Green function, keeping only the relevant terms:
\def\Am#1#2#3{\widetilde A_{#1}^{#2}(#3)}
\def\A#1#2#3{A_{#1}^{#2}(#3)}
\def\C2#1#2{({\cal C}_2)_{#1}^{#2}}
\beq\label{OPEfield1}
T\left( \Am{\mu}{a}{-p} \Am{\nu}{b}{p}\right)= (c_0)^{a b}_{\mu \nu}(p) +\cdots  \eeq
\[ + (c_2)^{a b \mu' \nu'}_{\mu \nu a' b'}(p) \ 
:\A{\mu'}{a'}{0} \ \A{\nu'}{b'}{0}: +\cdots\] 
\beq\label{OPEfield2}
T\left( \Am{\mu}{a}{p_1} \Am{\nu}{b}{p_2} \Am{\rho}{c}{p_3} \right)= \
(d_0)^{a b c}_{\mu \nu \rho}(p_1,p_2,p_3)\cdots \eeq
\[ + (d_2)^{a b c \mu' \nu'}_{\mu \nu \rho a' b'}(p_1,p_2,p_3) 
:\A{\mu'}{a'}{0} \ \A{\nu'}{b'}{0}: +\cdots 
\]

The Wilson coefficients depend on the renormalization scale. The leading
coefficients, $c_0(\mu)$ and $d_0(\mu)$ are exactly the  perturbative estimate of
the considered Green functions which is known to three loops. The first
subleading coefficients  $c_2(\mu)$ and $d_2(\mu)$ are proportional to $1/\mu^2$
for dimensional reasons and they also depend logarithmically on the
renormalization scale. We have computed to the leading logarithms approximation
the anomalous dimensions of these coefficients and that of the  $A^2$ operator. 

We have then performed the following test of OPE: A simultaneous fit was first
done of the two-point Green function and of $\alpha_s$ deduced from the symmetric
three-point Green functions computed on the lattice. We have used the known
dependence of the Wilson coefficient on the scale but, at this stage, fitted two
different values for  $<A^2>$. We then check the consistency of the resulting
values of $<A^2>$.

\newcommand{\Lams}{\Lambda_{\overline{\rm MS}}}
 
\[
\langle A^2 \rangle_{prop} =(1.55(17) {\rm GeV})^2;\,\langle A^2 \rangle_{alpha}
 = (1.9(3) {\rm GeV})^2
\]
\[\frac{\sqrt{\langle A^2 \rangle_{alpha}}}{\sqrt{\langle A^2 \rangle_{prop}}} 
 \ = \ 1.21(18);\,  \Lams = 233(28) {\rm MeV}\] 

The two fitted values of $<A^2>$ agree within one standard deviation. This is  a
quite interesting result which tends to confirm the hypothesis that the $A^2$
condensate is at the origin of the power corrections. It is interesting to stress
that if we use only the two loops formulae for the leading coefficients the
fitted condensates differ by more than twenty standard deviations.  This is a
lesson for the use of OPE: if the leading coefficients are not expanded far
enough in perturbation\footnote{ The meaning of ``enough'' depends on the case.}
the OPE cannot be applied. 
  
  \subsection{OPE for the asymmetric three-gluon Green function}
  
  OPE is more tricky to apply to the asymmetric Green function because of the
  presence of a zero momentum gluon. To be short, this mixes up the distinction
  between hard and soft gluons which is at the basis of the OPE approach. The
  issue has been discussed in \cite{FP}. Strictly speaking OPE applied to this
  operator introduces other operators than the identity and the $A^2$ considered
  above. One needs to consider gluon to vacuum matrix elements of  operators with
  an odd number of gluon fields. 
  
  In order to be able to relate nevertheless the asymmetric  three-gluon Green
  function measured on the lattice to the $A^2$ condensate the authors of
  \cite{FP} have introduced a ``vacuum insertion hypothesis'' (or factorization
  hypothesis) consisting in taking the matrix elements  of  three-gluon operators
  as the products of matrix elements of one gluon  operators and of two-gluon
  operators. Then, the only  non trivial matrix element is again  $<A^2>$ and one
  can perform the same exercise as above. The  combined fit of the propagator
  and  the asymmetric  $\alpha_s$ deduced from the asymmetric three-gluon Green
  function leads to:

\[\langle A^2 \rangle_{prop} = \left( 1.39(14) {\rm GeV} \right)^2;\,
\langle A^2 \rangle_{alpha} = \left( 2.3(6) {\rm GeV} \right)^2\]
\[\frac{\sqrt{\langle A^2 \rangle_{alpha}}}{\sqrt{\langle A^2 \rangle_{prop}}} \ =
 \ 1.7(3);\,
 \Lambda_{\overline {\rm MS}} =  260(18) {\rm MeV}\]
 
The agreement between both estimates of  \newline $<A^2>$ is less
satisfactory than in the previous case, maybe due to the added assumption of
factorization or to the necessity of going to four loops for the dominant
coefficients.

As a conclusion, we would like to emphasize the role of lattices as a beautiful
tool to study the QCD vacuum and in particular the vacuum condensates and also to
test the practical applicability of OPE. Notice also \cite{herve} that there is a
good evidence for power corrections in the gluon Green functions computed with
dynamical quarks. 

\section{Acknowledgments}
This work was supported by the European Community's Human potential 
programme under HPRN-CT-2000-00145 Hadrons/LatticeQCD.

\end{document}